# Activités collectives et instrumentation : étude de pratiques dans l'enseignement supérieur

**Christine Michel — Sébastien George — Elise Garrot**

**Laboratoire LIESP, INSA-Lyon**

*21, avenue Jean Capelle*
*F-69621 Villeurbanne Cedex, France*
*{Christine.Michel, Sebastien.George, Elise.Garrot}@insa-lyon.fr*

RÉSUMÉ. *Cet article présente les résultats d'une étude de terrain (enquête d'usage) portant sur les caractéristiques des Situations d'Apprentissage Collectives Instrumentées (SACI) dans l'enseignement supérieur selon le point de vue des acteurs (concepteur et tuteur) qui ont participé à leur conception ou à leur utilisation avec des étudiants. Notre propos est de déterminer les formes que prennent les SACI (en termes de scénario, d'outils, de type d'activité…) et les démarches pédagogiques mises en pratique par les acteurs enseignants. Nous détaillons les résultats de notre étude, faisant principalement ressortir une démarche « artisanale » des acteurs, c'est-à-dire fondée sur une utilisation opportuniste et pragmatique des outils disponibles pour s'adapter aux contextes de formation.*

MOTS-CLÉS : *Analyse et évaluation des usages, EIAH, Apprentissage collectif, Outil de travail collaboratif/coopératif.*

ABSTRACT. *This article presents the results of a study which concerns Instrumented Collective Learning Situations (ICLS) used in higher education and according to different actors's point of view (instructional designer and tutor). Considered actors have been involved in ICLS conception or in their use with students. We determine several forms of ICLS (in terms of scenario, tools, kind of activity) and what educational approaches are adopted by educational actors in their practices. We detail the results of our study, mainly by highlighting the "home-made" approach of the actors, that is based on an opportunist and pragmatic use of available tools in order to fit into the educational contexts.*

**Introduction**

Les travaux de recherche présentés dans cet article s'inscrivent dans le projet ACTEURS TICE (Activités Collectives et Tutorat dans l'Enseignement Universitaire : Réalités, Scénarios et usages des TICE) regroupant une équipe internationale et interdisciplinaire (STIC et SHS) de chercheurs. Ce projet a pour objectif général d'analyser et de comprendre les pratiques des acteurs – apprenants, tuteurs ou concepteurs (enseignants-auteurs) – impliqués dans le déroulement ou la conception de dispositifs de formation collectifs et instrumentés. Dans le cadre de ce projet nous nous intéressons plus spécifiquement à l'étude des Situations d'Apprentissage Collectives Instrumentées (SACI) dans l'enseignement supérieur (Bourriquen *et al.*, 2006). En effet, l'apprentissage collectif est utilisé pour favoriser l'apprentissage individuel à partir d'interactions entre apprenants (Dillenbourg, 1999). Dans ce contexte, des activités variées sont mises en place, telles que l'apprentissage par projet, les études de cas ou les jeux d'entreprise. Elles sont de plus en plus souvent réalisées sous une forme instrumentée pour pouvoir être utilisées de manière plus flexible (en particulier à distance). Parallèlement, des pressions extérieures (académique, économique, sociale) encouragent l'utilisation des Technologies de l'Information et de la Communication (TIC). En considérant l'importance actuelle des TIC en éducation et leur usage dans les activités collectives, nous avons décidé d'observer ce que nous nommons les SACI dans le contexte de l'enseignement supérieur. Nous tenons néanmoins à préciser que la finalité de cette recherche n'est pas de comprendre les mécanismes collectifs pour améliorer la conception de système d'apprentissage comme c'est le cas dans le champ des CSCL (*Computer Support for Collaborative Learning*) (Bannon, 1989) (ISLS, 2006) (Jermann, 2001).

Notre propos est de montrer, dans une démarche globale et interdisciplinaire, (1) les formes que peuvent prendre les SACI et leur intégration dans les programmes de formation, (2) le rôle des acteurs des SACI et leur ressenti par rapport à ce rôle, et (3) l'intentionnalité exprimée *a priori* par les concepteurs sous la forme de scénarios. Cette étude est réalisée par une observation et caractérisation des situations effectives sur le terrain. Après avoir défini plus précisément le terme SACI et la méthodologie d'observation, nous présentons les résultats obtenus.

**Problématique**

Nous définissons une SACI comme *une situation pédagogique avec un objectif d'apprentissage (de connaissances et/ou de compétences), des acteurs identifiés, une durée et un mode d'évaluation des apprenants, situation qui prend la forme d'une unité d'apprentissage scénarisée dans laquelle la production individuelle* *et/ou collective attendue est liée à une activité collective instrumentée par des artefacts informatiques* (Bourriquen *et al.*, 2006). Dans cette définition, le terme « collectif » est employé pour désigner à la fois les concepts « collaboratif » et « coopératif » (George, 2001), les sous-buts étant identiques dans une activité collaborative alors qu'il y a un partage de tâches entre acteurs dans une activité coopérative. Le terme « production » désigne tout résultat tangible, matérialisé et observable *a posteriori* d'une activité. Par exemple la production peut être une synthèse individuelle ou collective, des discussions provenant d'outils de communication ou de diffusion (forums, mails, blogs, chats), des traces de connexion, etc. Il est à noter qu'une SACI peut avoir différents niveaux de granularité – elle peut ainsi être une séance, une séquence ou une unité d'enseignement – dès lors qu'elle est délimitée par les caractéristiques énoncées dans sa définition (sa durée, son objectif d'apprentissage…).

Les plates-formes éducatives proposent de nombreux outils : *outils de communication* (chat, mail, forum, vidéoconférence, audioconférence), *outils de production et de partage* (partage d'applications, tableau blanc, éditeur de texte partagé, wiki, blog, portfolio) et *outils de gestion du travail collaboratif* (gestion de document, outil de planification et de coordination). Mais West *et al.* (2006), s'appuyant sur le modèle de Rogers (2003) pour comprendre le processus d'adoption de la technologie du point de vue de l'individu, montrent qu'il existe de nombreuses conditions à l'adoption de plates-formes informatiques pour qu'elles soient utilisées par les enseignants. Mahdizadeh *et al.* (2007), identifient 5 familles de critères pour expliquer l'usage des plates-formes : l'utilisation préalable de plates-formes, la perception de la valeur ajoutée de ces plates-formes pour l'apprentissage, l'opinion personnelle à propos des activités de type Web et des activités d'apprentissage instrumenté. L'adoption de la technologie est une étape, le développement de ressources nécessaires à leur propre action au sens de Beguin (2007) en est une autre. Ainsi, Ollagnier (2004) mentionne qu'« *être capable de bricoler des outillages pédagogiques figure sans aucun doute parmi les compétences des plus essentielles dans le métier. Appliquer à la lettre un outil existant ou un manuel ne marche pas. C'est la démarche pédagogique plus que l'outil brut qui va conditionner les apprentissages. Donc, c'est toute la subjectivité du formateur qui va, par le biais d'innombrables repères …/… lui faire prendre la « bonne » décision de l'attitude à avoir* ».

Nous considérons dans nos travaux les démarches pédagogiques de deux acteurs enseignants que sont : le concepteur pédagogique des situations d'apprentissage (scénarios, activités, contenus) et le tuteur qui encadre les apprenants durant le déroulement de ces situations. En effet, Mc Pherson & Nunes (2004) insistent autant sur l'importance du rôle du tuteur que sur celui du concepteur : *"the focus is frequently placed on design and developing information and communication technology (ICT) based environments and insufficient attention is given to the delivery process."*. Cette séparation des rôles nécessite d'une part l'écriture de recommandations de la part du concepteur et d'autre part la compréhension et le suivi de celles-ci par les tuteurs. Cependant, Casey & Greller (2005) observent que les tuteurs essaient d'adapter les contenus scénarisés à leurs étudiants sans aide, selon leur propre expérience.

Dans le cadre des SACI, nous sommes amené à nous demander s'il existe des démarches pédagogiques caractérisables ? Nous avons fait le choix de répondre, dans un premier temps à cette question, en observant les usages des outils TIC. Nous observerons en particulier si la mise à disposition de ces outils aux acteurs (tuteurs et apprenants) implique leur utilisation, et si leur utilisation implique une réelle activité collective entre apprenants. Nous observons si les concepteurs formalisent, lors des phases de conception, les démarches pédagogiques des activités collectives, et dans l'affirmative si les tuteurs et apprenants les respectent. Plus spécifiquement nous tentons de voir s'il existe des phénomènes de genèse instrumentale (Rabardel, 1995) composée de deux processus, en interaction, d'instrumentation et d'instrumentalisation. Au cours du processus d'instrumentation, ce sont les schèmes d'usage de l'utilisateur qui évoluent, se transforment, sont créés, s'incorporent aux schèmes déjà existants. Au cours du processus d'instrumentalisation, c'est l'artefact qui évolue avec l'émergence de nouvelles propriétés fonctionnelles. Nous tenterons d'expliquer ces phénomènes qui font que l'activité réalisée est différente de celle demandée (Beguin, 2007).

Nous sommes conscients qu'il est ardu d'explorer le réel effectif de ces pratiques pour en induire des interprétations et connaissances du comportement des individus et nous savons pertinemment que cette méthodologie d'analyse par l'observation des usages est lacunaire. En effet, comme dans tout processus de mesure, l'observation ne donne qu'une vision partielle de la réalité. De plus, comme le souligne Clot (2005), l'action réalisée et observable n'a pas le monopole du réel de l'activité. Le non réalisé, possible ou impossible, en fait aussi partie. Dans le contexte de ce projet nous ne sommes pas en mesure d'agir selon les recommandations d'analyses auto-réflexives de l'acteur des cliniques de l'activité (Clot, 2004), mais nous tenterons néanmoins d'adopter des méthodes d'observation dans lesquels différents points de vues sur les usages, ceux des tuteurs et des concepteurs en particulier, seront simultanément pris en compte.

Ainsi, notre travail vise à observer :

– dans quelle proportion les SACI, telles que nous les définissons, existent réellement,

– la réalité des pratiques des acteurs afin d'évaluer les éventuels écarts entre les recommandations proposées par le monde de la recherche et les usages effectifs,

– les écarts existants entre les activités prescrites par les concepteurs pédagogiques et le déroulement effectif de celles-ci,

Un objectif de cette recherche est de formuler des hypothèses pour expliquer les différences observées au niveau des deux derniers points précédents et d'ouvrir des pistes de réflexion pour faciliter la mise en place et la réalisation de SACI.

**Méthodologie**

L'observation a été réalisée à partir d'entretiens avec des enseignants, entretiens guidés par un questionnaire. Ce questionnaire final a été construit selon une démarche itérative et participative (Mackay *et al.*, 1997). Cette dernière consiste en effet à intégrer les utilisateurs au processus de conception et d'évaluation des livrables et ainsi d'en valider avec eux la pertinence. Ainsi, un premier guide d'entretien et un questionnaire préalable ont servi à la construction du questionnaire final.

*Echantillon d'étude*

Un questionnaire préalable et un entretien semi-directif ont été réalisés avec un concepteur du campus numérique FORSE (FORSE, 2006) et deux tuteurs du campus Spiral (SPIRAL, 2006). Les résultats ont permis d'affiner la définition d'une SACI en précisant ses caractéristiques spécifiques comme le niveau de granularité (séance, séquence, unité d'enseignement, etc.) et d'identifier, parmi les 13 terrains d'observation (définis dans le montage du projet ACTEURS TICE), ceux intégrant des SACI dans leur dispositif de formation et donc intéressant à observer plus en détail.

Les formations observées dispensent des cours à distance, en présence ou de manière hybride ; concernent différentes disciplines (sciences du langage, sciences de l'éducation, informatique, management…) et sont localisées en France, en Suisse et au Canada (Québec). Nous pensions pouvoir identifier 2 ou 3 SACI par terrain partenaire du projet ACTEURS TICE et choisir notre échantillon d'étude en panachant les disciplines et les grandes orientations pédagogiques choisies. En fait, cette enquête préalable a montré que les SACI n'étaient pas aussi développées dans le milieu de l'enseignement supérieur que nous l'avions pensé, seules 13 SACI ont été en effet identifiées comme correspondant à notre définition d'une SACI. Elles sont présentées dans le tableau ci-dessous (Tableau 1) et seront décrites plus précisément, via une enquête, par les acteurs qui ont participé à leur conception et à leur utilisation avec des étudiants.

Dans le Tableau 1, la part de présentiel et le nombre de sessions d'une SACI depuis sa création (i.e. son ancienneté) sont déterminés par les acteurs que nous avons interviewés.

| Num SACI | Nom SACI | Terrain | type activité | part presentiel (en %) | nb sessions | Rôle personnes interviewées | Num obs |
|---|---|---|---|---|---|---|---|
| 1 | TP anglais | Chambéry IUT | débat | 100 | 3 | Concepteur | 1 |
| | | | | | | Tuteur | 2 |
| 2 | Cours 3D | VCIEL | résolution problème | 0 | 1 | Tuteur | 3 |
| 3 | Cours php | VCIEL | résolution problème | 0 | 1 | Tuteur | 4 |
| 4 | Génie logiciel | Teluq | projet | 0 | 9 | Concepteur | 5 |
| | | | | | | Tuteur | 6 |
| 5 | UE conception multimédia | VCIEL | résolution problème | 0 | 1 | Concepteur et tuteur | 7 |
| | | | | | | Concepteur et tuteur | 8 |
| 6 | Projet info S2 : conduite d'un projet informatique | FISAD | projet | 36 | 6 | Concepteur et tuteur | 9 |
| 7 | jeu d'entreprise | IUT-B | projet | 50 | 1 | Concepteur et tuteur | 10 |
| 8 | DESS UTICEF : bases conceptuelles | Tecfa (UTICEF) | recherche info - débat | 0 | 12 | Concepteur | 11 |
| 9 | Master 1 Education et nouvelle technologie | FORSE | étude de cas | 5 | 2 | Tuteur | 12 |
| | | | | | | Concepteur | 13 |
| 10 | JIFE (Jeu d'initiation et de formation à l'entreprise) | E-Miage | projet | 30 | 3 | Concepteur et tuteur | 14 |
| 11 | Licence sciences de l'Education FOAD | FORSE | étude de cas - recherche info | 0 | 6 | Tuteur | 15 |
| | | | | | | Concepteur | 16 |
| 12 | TP electronique avec tablet-PC | ECL-EAT | étude de cas | 100 | 2 | Concepteur et tuteur | 17 |
| 13 | TP automatique avec tablet-PC | ECL-EAT | débat - résolution problème | 100 | 2 | Concepteur et tuteur | 18 |

**Tableau 1.** *Récapitulatif des 13 SACI étudiées.*

*Modalité d'enquête*

Le questionnaire final comporte 207 questions. Il a été construit pour recueillir des données sur la conception, la mise en place et le déroulement d'une SACI tant au niveau de la description du contexte global dans lequel s'inscrit la SACI (contenu, public, diplôme, discipline, genèse de la SACI), que des usages des outils (attendus ou effectifs) par les concepteurs, tuteurs et apprenants, et plus spécifiquement, sur le type d'instrumentation et les modalités d'interactions collectives mises en œuvre dans la SACI.

Comme indiqué précédemment, le questionnaire a été construit en deux temps. Dans une première phase, nous avons établi un questionnaire de 69 questions relatives au contexte et objectifs pédagogiques d'une SACI, à sa description technique et à l'usage que pouvaient en faire les acteurs (concepteur, tuteur, apprenants). Parallèlement nous avons défini, dans un tableau à double entrée, comment chaque question pouvait renseigner différents indicateurs pour décrire des SACI répartis selon 5 axes (une croix à la jonction de la ligne et de la colonne indique que la question renseigne l'indicateur) :

– *Axe Contexte* : intégration, discipline, maturité, etc.

– *Axe Acteurs* : concepteur de la SACI (expérience d'utilisation des outils, de la conception pédagogique, etc.), Tuteur (expérience d'utilisation des outils, du tutorat, etc.), Apprenant (expérience d'utilisation des outils, du travail collectif, etc.).

– *Axe Pédagogie* : objectifs pédagogiques, évaluation des apprentissages, évaluation du travail collectif, etc.

– *Axe Travail collectif* : types de tâches, part individuel/collectif, part synchrone/asynchrone, etc.

– *Axe Technologie* : environnements concepteur/tuteur/apprenant (type d'outils, flexibilité, contraintes, etc.),

Ce mode d'organisation avec le tableau nous a permis non seulement de tester la pertinence des questions et le degré de complétude du questionnaire, mais aussi d'aider à l'analyse des données recueillies : évaluation synthétique du positionnement des SACI dans chaque axe, caractères invariants ou critères distinctifs, degrés de dépendance entre les indicateurs. De plus cette analyse avait pour vocation de faire ressortir l'existence de liens entre les objectifs pédagogiques et les outils utilisés. Dans les faits, ce tableau n'a pas servi à révéler ces liens car les situations étaient trop différentes en termes d'objectifs pédagogiques. Cela pourra servir pour une future recherche avec davantage de SACI étudiées. Le tableau a en revanche bien rempli son rôle de contrôle et test du questionnaire et a permis d'identifier un certain nombre de dysfonctionnements dans la formulation et l'étendue des questions. Ainsi, le questionnaire final utilisé dans l'étude a été complété pour arriver à 207 questions. Elles ont été regroupées en trois grandes parties renseignant :

– La *définition* de la SACI :

- informations générales sur la SACI (part de présentiel, discipline, usage de l'outil informatique, motivation pour la mise en place, nombre de sessions depuis la création, formation diplômante) ;

- renseignements sur la conception et l'usage de la SACI (durée de la session, objectifs pédagogiques, type de travail coopératif et/ou collaboratif, objectif de la collaboration/coopération, proportion de travail individuel/collectif, présence d'un scénario pédagogique explicite, définition de rôles pour les apprenants et les enseignants, description de l'évaluation des apprentissages, etc.).

– La *caractérisation des acteurs impliqués* dans les SACI :

- les apprenants (nombre d'apprenants de la session, niveau de connaissances relatives à la discipline enseignée, niveau de pratique des outils informatiques utilisés, niveau de pratique du travail collectif, nombre moyen d'apprenants par groupe collectif, critères de constitution des groupes, nombre de tuteur par apprenant, etc.) ;

- les tuteurs (nombre, ont-ils participé à la conception, enseignent-ils habituellement en présentiel, niveau de pratique des outils informatiques utilisés, niveau de pratique du tutorat à distance, ont-ils eu une formation au tutorat, fonctions attribuées dans la SACI) ;

- les concepteurs (nombre, niveau d'expertise dans la conception d'activités pédagogiques de type SACI, niveau de spécialisation dans le contenu, niveau de pratique des outils informatiques utilisés, ont-ils eu une formation à la conception de SACI ?)

– Les *caractéristiques de l'environnement informatisé* supportant la SACI :

- description des outils de communication, de production/ diffusion/partage, de gestion du travail collectif proposés et utilisés (par les apprenants et par les tuteurs) ;

- description des outils de conception pour les concepteurs, description et usage des ressources pédagogiques, possibilité de « contrôle » du déroulement du scénario pédagogique, aide/contrainte/contournement/ ressentie par l'usage des outils, évolutivité possible, attentes.

Ce questionnaire a été formalisé sous le logiciel Sphinx. L'ensemble des 207 questions n'étant pas adressé à tous les acteurs, nous avons construit 3 sous questionnaires ; un pour chaque type d'acteur.

*Recueil des données*

Nous avons réalisé avec le questionnaire SACI 18 entretiens, entre avril et novembre 2006, auprès des acteurs des 13 SACI présentées précédemment (voir Tableau 1). Nous avons délibérément choisi de collecter ces informations par le biais d'entretiens plutôt que de mettre en ligne le questionnaire. En effet, les situations pédagogiques de type SACI peuvent être considérés comme des situations d'innovation dans lesquelles le vocabulaire et les concepts ne sont pas stabilisés et les processus ou interactions pas toujours bien formalisés par les acteurs. Ce phénomène est amplifié si l'on considère que les acteurs concernés en ont des perceptions différentes. Il est donc nécessaire de discuter, préciser, approfondir ceux-ci lors des entretiens.

Nous avons choisi de n'interroger que les concepteurs et les tuteurs ayant participé à la SACI et de ne pas interroger les apprenants. Le questionnaire n'ayant pas été élaboré pour être rempli de manière autonome et nécessitant un entretien pour recueillir des informations riches, nous avons préféré nous concentrer sur les acteurs enseignants que nous pouvions rencontrer. La plupart des SACI étant terminées au moment de notre enquête, il était difficile d'avoir suffisamment d'étudiants, condition nécessaire pour avoir une bonne représentation des activités des étudiants. Néanmoins, des informations sur ces activités seront présentes dans notre enquête par les points de vue des acteurs enseignants.

Pour chaque SACI, nous avons recueilli, lorsque cela était possible, le témoignage d'un concepteur et d'un tuteur. Cependant, dans certains cas, le même acteur était concepteur et tuteur. Chaque entretien a duré une heure environ par acteur, en face-à-face pour la plupart, par téléphone ou par visioconférence (pour le Canada et la Suisse). Lors de la prise de rendez-vous par e-mail avec l'enquêté, nous joignions le questionnaire. L'enquêté avait ainsi la possibilité de prendre connaissance du questionnaire avant l'entretien, et l'avait en main lors de l'entretien. Deux enquêteurs de notre équipe étaient présents, un seul posant les questions, les deux notant les réponses, afin de recueillir le maximum de données qualitatives pour les réponses ouvertes et pour les commentaires associés aux réponses (explication de la réponse choisie en cas d'hésitation, clarification du vocabulaire utilisé…). L'ensemble des réponses a ensuite été saisi sous le logiciel Sphinx pour en faciliter le traitement. Lors de la saisie, les réponses aux questions ouvertes notées par les deux enquêteurs d'un même entretien étaient mises en commun.

*Analyse et traitement des données*

Le processus d'analyse est quantitatif pour les questions fermées (analyse statistique de type dénombrement, moyenne, …, sur des tableaux unidimensionnels ou croisés). Les questions ouvertes ont été traitées dans un second temps de manière qualitative pour affiner les résultats et enrichir l'analyse.

Les questions fermées ont été analysées automatiquement (comptage, fréquence, moyenne) et manuellement (comparaison, croisement d'informations, recodification, …). Nous avons ainsi construit, principalement, à l'aide d'Excel, des tableaux comparatifs ou croisés entre l'utilisation *prescrite* de la SACI et l'usage *effectif* pendant le déroulement de la SACI. Ces tableaux ont porté sur les prescriptions des concepteurs envers les tuteurs versus l'usage effectif des tuteurs, ainsi que les prescriptions des tuteurs versus l'usage effectif des apprenants. En terme d'usage, nous avons principalement observé la fréquence et la facilité d'utilisation des outils informatiques de communication, de production/diffusion/partage et de gestion du travail collectif ainsi que la formalisation et le respect du scénario original.

Les questions ouvertes ont été analysées selon des techniques classiques d'analyse de contenu (i.e. sans utilisation de logiciel d'analyse textuelle) en traitant les réponses individuellement pour chaque SACI. Le but était double. Il s'agissait d'une part, d'illustrer les résultats qualitatifs obtenus de façon par exemple à préciser le contexte d'utilisation des outils, et d'autre part, d'apporter des éléments d'interprétation de résultats quantitatifs spécifiques ou intéressants. Nous nous sommes particulièrement intéressés aux données concernant les scénarios des SACI : leur précision, leur mise en place effective et leur déroulement. Cette analyse a notamment permis d'interpréter les observations sur l'usage effectif des outils par rapport à l'usage prévu.

**Résultats**

Avant de présenter les résultats, nous précisons que certaines données sont présentées en pourcentage même si l'échantillon est petit (18 observations), ceci dans un souci de clarté de lecture des résultats.

*Profil global des SACI*

Les premiers résultats sur le profil global des SACI observées montrent qu'elles s'inscrivent toutes dans des formations diplômantes. L'échantillon étudié contenait une bonne diversité de disciplines (anglais, informatique, infographie, ingénierie de la formation à distance, marketing, comptabilité, gestion d'entreprise, électronique, automatique). Les *motivations* qui ont poussé à la création de ces SACI sont :

– d'agir sur une formation existante en présence (resp. à distance) en la complétant (35% resp. 23%) ou en la remplaçant/améliorant (47% resp. 11,8%),

– de créer une nouvelle formation (41,2%)

– de chercher de nouveaux publics (35,3%).

Les expériences sont assez jeunes puisque 40% d'entre elles en sont à leur $1^{ere}$ session et que 70% ont été réalisées sur moins de 4 sessions. La *granularité* est très variable. Nous trouvons autant de SACI sur des périodes courtes (2h à, 2 jours) que sur des périodes très longues (1 semestre à 1 an). La part d'*enseignement en présence* est faible (inférieure à 10% du temps d'apprentissage dans environ 60% des cas). Les *objectifs* sont assez variés allant de réalisation de tâches simples (apprendre du vocabulaire, apprendre la manipulation d'un logiciel de conception graphique), à des tâches plus évoluées d'analyse globale et de gestion (de ressources, de projet, d'hommes).

*Description des activités d'apprentissage et outils de communication*

*Activités d'apprentissage*

Il n'y a pas de *type d'activité* privilégié (voir tableau 2) puisque nous relevons approximativement autant de situations d'apprentissage concernant de la recherche d'informations, du débat, de l'écriture collaborative, du projet, de l'étude de cas et de la résolution de problèmes. Les activités sont même souvent combinées : écriture collaborative dans un projet, recherche d'information et résolution de problème, débat et écriture collaborative par exemple.

| % | Recherche d'information | Débat | Lecture/écriture | Projet | Etude de cas | Résolution de problème |
|---|---|---|---|---|---|---|
| **Non ou plutôt non** | 50 | 38,9 | 33,3 | 44,4 | 44,4 | 50 |
| **Oui ou plutôt oui** | 50 | 61,1 | 66,7 | 55,6 | 55,6 | 50 |

**Tableau 2.** *Types d'activités proposées aux apprenants (en % des SACI observées)*

*Coopératif versus collaboratif*

Le temps de travail *collectif* prévu par le concepteur est en moyenne de 64,38% par SACI (le reste étant du travail *individuel*), ce qui est généralement suivi dans la pratique (60,71%). Le travail est plutôt de nature *coopérative* que *collaborative* (60% en effectif) mais ce chiffre est à prendre avec précaution. Il semblerait que les acteurs aient eu du mal à distinguer la collaboration de la coopération, même si une définition était donnée lors des entretiens. En effet, nous avons observé par exemple des réponses contradictoires pour une même SACI (le concepteur d'une SACI indiquant une proportion d'activités collaboratives de 100% et le tuteur de 0%).

*Activités synchrones et asynchrones*

Le récapitulatif des résultats concernant les volumes d'activité est disponible dans (Michel *et al.,* 2007a) (Michel *et al.,* 2007b). Nous y observons que globalement, les activités *synchrones* sont très souvent prescrites par le concepteur pour les apprenants (87,5%) et pour les tuteurs (100%). Si les activités *asynchrones* sont un peu moins prescrites (avec respectivement 68,75% et 64,29%) elles sont bien suivies (seulement 6,25% des acteurs les utilisent « pas souvent ») alors que les activités *synchrones* le sont moins (31,25% les utilisent « pas souvent »).

Le respect de la prescription de l'activité est donc très variable. Plus précisément, les figures 1 à 4 montrent le nombre (entre « pas du tout » et « souvent » sur une échelle de 5) d'activité synchrone ou asynchrone *prévue* (visible par un tiret) et *effectuée* (visible par un rond) par les apprenants et les tuteurs dans chaque SACI. Les SACI sont numérotées de 1 à 18 en abscisse. Les écarts entre les activités prescrites et réalisées sont identifiés par un trait de jonction. Les traits noirs indiquent des « sous-utilisations », et les clairs indiquent des « sur-utilisations », par rapport à ce qui était prévu. Les figures 1 et 2 montrent que les activités *asynchrones* ont globalement été bien prévues pour les apprenants et les tuteurs alors que les figures 3 et 4 montrent qu'elles ne l'ont pas forcement été pour les activités *synchrones*.

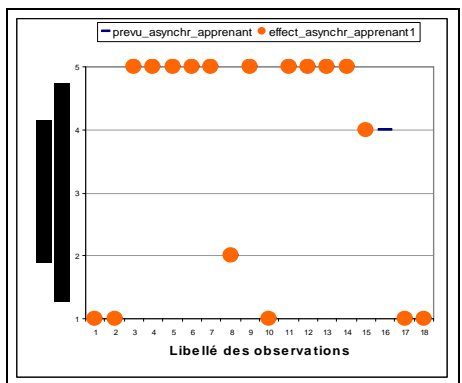

**Figure 1.** *Comparatif des activités asynchrones des apprenants prévues/effectives*

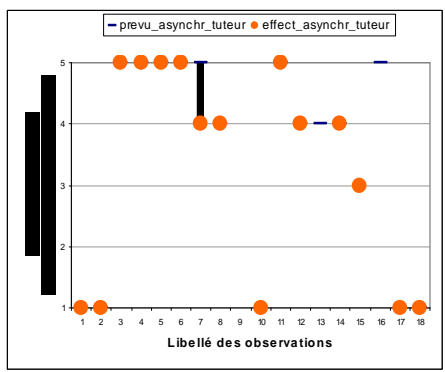

**Figure 2.** *Comparatif des activités asynchrones des tuteurs prévues/effectives*

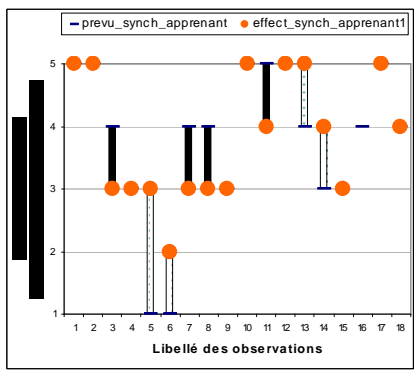

**Figure 3.** *Comparatif des activités synchrones des apprenants prévues/effectives*

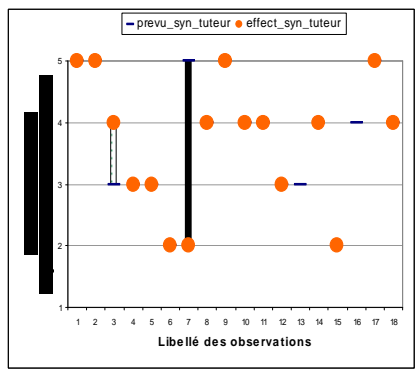

**Figure 4.** *Comparatif des activités synchrones des tuteurs prévues/effectives*

Dans les figures 3 et 4, des sous-utilisations d'outils synchrones sont souvent identifiées chez les apprenants pour les observations 3, 7, 8 (SACI dans VCIEL) et 11 (SACI dans UTICEF) et ponctuellement chez les tuteurs (observation 7). Dans le cas de VCIEL, les tuteurs et concepteurs interrogés ont observé que les séances régulières de chat prévues dans le scénario n'avaient pas été suivies par les apprenants qui préféraient contacter leurs tuteurs via le forum (en l'utilisant comme un mail et sans lire les messages préalablement postés) alors qu'ils ouvraient parallèlement des conversations avec les autres apprenants via des outils de communication instantanée (de type MSN ou Yahoo Messenger) en dehors de la SACI. Dans le cas de l'observation 11, il n'y a pas de remarque particulière du tuteur qui nous permette une interprétation. Des sur-utilisations d'outils synchrones par les apprenants sont identifiées pour les observations 5, 6 (SACI dans Teluq), 13 (SACI dans FORSE) et 14 (SACI dans E-miage). Dans le cas de la Teluq, l'activité des apprenants est prévue pour être totalement asynchrone : le chat n'est pas proposé mais les apprenants ont la possibilité de contacter le tuteur par téléphone. C'est ce type d'usage qui a provoqué la sur-utilisation. Dans le cas de FORSE, elle s'explique par une confusion d'usage par les apprenants qui utilisent le forum comme messagerie et la messagerie comme forum. Les questions ouvertes de l'observation 14 n'étant pas renseignées, nous ne pouvons l'interpréter.

*Outils de travail collectif*

Outils de communication

Les résultats, publiés dans (Michel *et al.,* 2007a), montrent que les outils de communication les plus fournis sur les plates-formes sont le mail (68,75% pour les apprenants et 76,47% pour les tuteurs), le forum (68,75% pour les apprenants et 64,7% pour les tuteurs) et le chat (56,25% pour les apprenants et 58,82% pour les tuteurs). De manière plus détaillée, si nous observons à présent l'usage, le tableau 3 montre que ces trois outils sont assez souvent utilisés lorsqu'ils sont proposés. Nous observons que les apprenants utilisent plus les forums que le mail alors que les tuteurs préfèrent le mail. Si nous rapprochons cette observation de la confusion d'usage identifiée ci-dessus, nous pouvons convenir que le caractère asynchrone de ces échanges est primordial alors que le caractère privé et personnel de l'email *vs.* public et collaboratif du forum ne l'est pas et que les apprenants utilisent presque indifféremment l'un ou l'autre. Les chats sont aussi utilisés quand ils ne sont pas proposés. Cet usage est vraisemblablement lié aux chats externes, ouverts par les apprenants en dehors de la SACI pour des raisons de confidentialité ou d'utilisabilité.

| Usage des apprenants | Proposé et non utilisé | Proposé et un peu utilisé | Proposé et souvent utilisé | Non proposé et non utilisé | Non proposé et un peu utilisé | Non proposé et souvent utilisé |
|---|---|---|---|---|---|---|
| e-mail | 0 | 37,5 | 31,25 | 31,25 | 0 | 0 |
| forum | 0 | 18,75 | 50 | 31,25 | 0 | 0 |
| Chat | 0 | 31,25 | 25 | 31,25 | 12,5 | 0 |
| Visio conference | 0 | 12,5 | 0 | 81,25 | 0 | 6,25 |
| Audio conference | 0 | 0 | 0 | 100 | 0 | 0 |
| Téléphone | 0 | 12,5 | 0 | 81,25 | 6,25 | 0 |
| Courrier | 6,25 | 6,25 | 0 | 81,25 | 6,25 | 0 |
| Face a face | 0 | 6,67 | 33,33 | 60 | 0 | 0 |
| Usage des tuteurs | Proposé et non utilisé | Proposé et un peu utilisé | Proposé et souvent utilisé | Non proposé et non utilisé | Non proposé et un peu utilisé | Non proposé et souvent utilisé |
| e-mail | 0 | 35,29 | 41,18 | 23,53 | 0 | 0 |
| forum | 5,88 | 23,53 | 35,29 | 35,29 | 0 | 0 |
| Chat | 5,88 | 29,41 | 23,53 | 29,41 | 11,76 | 0 |
| Visio conference | 0 | 17,65 | 0 | 76,47 | 5,88 | 0 |
| Audio conference | 0 | 11,76 | 0 | 88,24 | 0 | 0 |
| Téléphone | 0 | 35,29 | 11,76 | 52,94 | 0 | 0 |
| Courrier | 11,76 | 5,88 | 0 | 76,47 | 5,88 | 0 |
| Face a face | 0 | 21,43 | 35,71 | 42,86 | 0 | 0 |

**Tableau 3.** *Outils de communication (en % des SACI observées)*

Outils de partage et de production de ressources

La plupart du temps, les SACI observées proposent rarement des *outils de partage* et de *production* de ressources (tableau 4). Lorsqu'ils sont proposés, nous retrouvons principalement des éditeurs de texte partagé et des tableaux blancs qui sont plutôt bien utilisés par les apprenants et les tuteurs. À la lecture des questions ouvertes sur les scénarios, nous avons pu constater que ces outils étaient souvent intimement liés à la réalisation d'une activité spécifique dans la SACI. Cette conclusion est corroborée par le fait que les tuteurs et les apprenants n'utilisent pas ces outils quand ils ne sont pas prescrits et qu'ils ne les utilisent pas indépendamment de l'activité pédagogique pour s'organiser. Les wikis sont proposés dans 50% des cas aux tuteurs mais ne sont jamais utilisés ; les concepteurs ne précisent d'ailleurs pas, dans le scénario, la raison pour laquelle les tuteurs doivent l'utiliser.

| Usage des apprenants | Proposé et non utilisé | Proposé et un peu utilisé | Proposé et souvent utilisé | Non proposé et non utilisé | Non proposé et un peu utilisé | Non proposé et souvent utilisé |
|---|---|---|---|---|---|---|
| **Partage d'application** | 0 | 6,25 | 0 | **87,5** | 6,25 | 0 |
| **Tableau blanc** | 0 | **12,5** | **12,5** | **75** | 0 | 0 |
| **Editeur de texte partagé** | 6,25 | 0 | **25** | **68,75** | 0 | 0 |
| **Wiki** | 12,5 | 0 | 0 | **87,5** | 0 | 0 |
| **Blog** | 6,25 | 0 | 0 | **93,75** | 0 | 0 |
| **Portfolio** | 0 | 0 | 0 | **100** | 0 | 0 |
| Usage des tuteurs | Proposé et non utilisé | Proposé et un peu utilisé | Proposé et souvent utilisé | Non proposé et non utilisé | Non proposé et un peu utilisé | Non proposé et souvent utilisé |
| **Partage d'application** | 0 | 12,5 | 0 | 87,5 | 0 | 0 |
| **Tableau blanc** | 0 | **6,67** | **6,67** | 86,66 | 0 | 0 |
| **Editeur de texte partagé** | 0 | **6,25** | **12,5** | 81,25 | 0 | 0 |
| **Wiki** | **50** | 0 | 0 | 50 | 0 | 0 |
| **Blog** | 6,25 | 0 | 0 | 93,75 | 0 | 0 |

**Tableau 4.** *Outils de partage et de production de ressources (en % des SACI observées)*

Outils de gestion du travail collectif

Les outils de *gestion du travail collectif* sont assez peu proposés. Quand ils le sont, c'est sous la forme d'outils de gestion de document, d'outils spécifiques et d'outils de planning et dans une faible mesure d'outils d'*awareness* (pour favoriser la conscience des activités du groupe). Les apprenants utilisent ces outils de manière irrégulière lorsqu'ils sont prescrits, sauf pour les outils de planning. Les tuteurs respectent plutôt bien les prescriptions (très peu déclarent ne jamais les utiliser lorsqu'ils sont prescrits). Lorsqu'ils ne sont pas prescrits, ils ne sont pas utilisés sauf par les tuteurs pour les outils d'*awareness* et les outils spécifiques. Nous pouvons supposer que les tuteurs détournent l'usage initial de ces outils pour satisfaire des besoins non pris en compte par le concepteur (comme l'évaluation du travail d'un élève et de son degré d'investissement grâce aux outils d'*awareness*, la composition d'un groupe et la collaboration au sein et entre les groupes grâce aux outils spécifiques d'organisation).

| Usage des apprenants | Proposé et non utilisé | Proposé et un peu utilisé | Proposé et souvent utilisé | Non proposé et non utilisé | Non proposé et un peu utilisé | Non proposé et souvent utilisé |
|---|---|---|---|---|---|---|
| **Agendas** | 0 | 0 | 26,67 | 73,33 | 0 | 0 |
| **Gestion de document** | 18,75 | 12,5 | 12,5 | 56,25 | 0 | 0 |
| **Awareness** | 6,67 | 0 | 6,67 | 86,66 | 0 | 0 |
| **Outils spécifiques** | 15,38 | 0 | 15,38 | 69,24 | 0 | 0 |
| Usage des tuteurs | Proposé et non utilisé | Proposé et un peu utilisé | Proposé et souvent utilisé | Non proposé et non utilisé | Non proposé et un peu utilisé | Non proposé et souvent utilisé |
| **Agendas** | 0 | 0 | 25 | 75 | 0 | 0 |
| **Gestion de document** | 0 | 20 | 13,33 | 66,67 | 0 | 0 |
| **Awareness** | 0 | 7,69 | 7,69 | 76,92 | 7,69 | 0 |
| **Outils spécifiques** | 7,14 | 7,14 | 7,14 | 71,43 | 7,14 | 0 |

**Tableau 5.** *Outils de gestion du travail collectif (en % des SACI observées)*

Autres ressources

Les SACI proposent d'autres ressources informationnelles sous forme de documentation papier (dans 42% des cas), de documentation en ligne (dans 64% des cas) et des ressources interactives (dans 20% des cas). Ces trois ressources sont massivement utilisées.

*Formalisation de l'activité*

La grande majorité des SACI ont été réalisées suivant un scénario, exceptées deux. Dans le tableau 6 ci-dessous, nous précisions le type et les outils de formalisation de ces scénarios. Pour chaque scénario, nous considérons s'il y a un modèle formalisé pour la scénarisation, des outils pour la conception, un scénario spécifique pour les interactions.

| Modèle de scénario | | | Outils pour la conception des scénarios | | | Interactions scénarisées | | |
|---|---|---|---|---|---|---|---|---|
| non | oui | sans rep. | non | oui | sans rep. | oui | non | sans rep. |
| 31 | 50 | 19 | 50 | 25 | 25 | 44 | **56** | 0 |

**Tableau 6.** *Formalisation de l'activité (en % des SACI avec scénario)*

Dans la grande majorité des cas, les SACI sont mises en place avec une intention pédagogique d'apprendre à collaborer ou de collaborer pour apprendre. Le fait d'observer que dans près de la moitié des cas les interactions ne sont pas scénarisées nous incite à penser que le niveau de description du scénario est très global dans ces cas là. Pour celles qui formalisent ces interactions, nous observons qu'elles ne le font pas de manière standardisée (i.e. sans utilisation de langage comme SCORM ou IMS-LD (IMS, 2007) (Koper, 2006). Concernant le modèle pédagogique, la plupart des SACI sont des adaptations de formations présentielles. Celles qui ne le sont pas s'appuient soit sur une formalisation des objectifs pédagogiques et connaissances (2 cas), soit sur un jeu commercial déjà scénarisé (1 cas). Les scénarios décrivent le plus souvent des alternances de réalisations de tâches individuelles (généralement lecture de cours ou réalisation de TD) et de tâches de communication durant lesquelles les apprenants peuvent solliciter de l'aide ou des informations complémentaires. Nous n'observons que très peu de réalisation *collaborative* de tâches entre apprenants.

La démarche de scénarisation est donc globalement très artisanale en ce qui concerne la conception des SACI (peu de formalisation et de standardisation). L'instrumentation est souvent utilisée pour pallier la distance et non pour créer des situations d'apprentissage innovantes et adaptées aux apprenants. Les outils apportent surtout des possibilités de communication durant l'apprentissage que de nouvelles formes de travail collectif. Ces conclusions sont cohérentes avec celles de Mahdizadeh *et al.* (2007) qui montrent que le e-learning dans l'enseignement supérieur aux États-Unis est encore à un stade d'usage peu avancé : il n'est pas bien intégré dans les formations et les enseignants n'utilisent que les possibilités élémentaires des outils. Il y a donc un écart fort entre les résultats des chercheurs en EIAH qui ont identifié de multiples possibilités d'apprentissages collectifs et les mises en œuvre effectives dans les situations *e-learning*. Ceci explique la tendance actuelle vers de nouveaux types de recherches qui s'intéressent de plus près aux pratiques réelles des enseignants. Nous faisons notamment référence aux travaux d'identification de pratiques effectives des enseignant, des études mettant en avant le concept de « bricolage » (Ollagnier, 2003) (Caron, 2007). D'autres travaux visent à favoriser l'échange entre enseignants de pratiques et d'expériences vécues (Garrot *et al.*, 2007).

**Conclusion**

Cette étude avait pour objet de voir dans quelle mesure les SACI (situations d'apprentissage collectives instrumentées) étaient effectivement mises en œuvre dans l'enseignement supérieur. Au travers d'une enquête, nous avons montré que la motivation poussant à la création de SACI est généralement de remplacer/améliorer une formation déjà existante en présence par une formation à distance ou hybride.

Cependant, ces types de formation restent principalement fondés sur l'apprentissage individuel et les SACI sont encore assez peu développées.

Dans notre enquête, les SACI observées sont assez jeunes, n'ont pas de traits caractéristiques forts en termes d'objectif d'apprentissage, granularité ou type d'activité mais dans l'ensemble, les activités prévues sont globalement bien réalisées par les apprenants et les tuteurs. Les outils de communication y sont massivement utilisés mais de manière détournée ou concourante. En effet, les outils asynchrones sont souvent utilisés mais nous observons des confusions d'usage entre le mail et le forum par les apprenants qui utilisent le forum comme messagerie et la messagerie comme forum. Concernant les outils synchrones, les étudiants ouvrent régulièrement des *chats* ou des sessions de messagerie instantanée en dehors des plates-formes institutionnelles. Cela nous laisse supposer (1) qu'ils préfèrent garder leurs outils habituels (2) qu'ils ont besoin d'avoir des espaces privés en dehors de tout contrôle (3) que les rendez-vous ou la finalité des séances de *chat*s prévues dans le scénario ne leurs conviennent pas. Les outils de partage et de production de ressources sont assez peu utilisés, ou alors pour accompagner ponctuellement une activité et non pour aider les acteurs à réaliser globalement leurs activités d'apprentissage ou d'enseignement. Ces détournements d'usage pourraient être étudiés de manière spécifique notamment en complétant notre travail par des enquêtes auprès des apprenants.

Concernant la formalisation des activités pédagogiques, nous avons pu observer que les scénarios d'apprentissage les plus usuels ne rentrent pas dans un grand niveau de précision, sont assez peu variés et ne sont pas décrits dans un formalisme particulier ou en suivant les recommandations d'un standard. Ils reposent en effet souvent sur une alternance de réalisations de tâches individuelles et de communication, avec très peu de réalisations réellement collaboratives entre apprenants. Cette étude montre que les possibilités de mise en œuvre des activités collectives instrumentées par les concepteurs et les tuteurs ne sont qu'assez peu exploitées ou explorées. Des études complémentaires seraient nécessaires pour déterminer si cela est dû à une réaction pragmatique liée au manque de moyens matériels, à une difficulté à mettre en application des recommandations issues de la recherche ou bien à une méconnaissance ou une mauvaise compréhension/interprétation de ces recommandations. En particulier, une étude plus précise des recommandations permettrait de déterminer si elles sont effectivement assez formelles et généralisables ou si elles résultent de conclusions partielles, tirées d'expériences spécifiques. Si nous considérons le processus de transfert technologique dans le contexte d'innovation développé par Perrin (1983), nous pouvons nous demander si l'enseignant développe des techniques nouvelles à partir de celles dont il dispose ou adapte, modifie et transforme les SACI pour les conformer à ses propres constructions de son activité.

Dans notre étude, nous n'avons pas décelé de problème dans l'acculturation technique, ce qui est parfois un réel frein. L'erreur consiste à penser que si l'usage du dispositif est maîtrisé, si les recommandations du monde de la recherche sont compréhensibles, alors les concepteurs et tuteurs sont à même d'imaginer des SACI spécifiquement adaptées, utiles et utilisables. Or, nous venons de voir que ce n'est pas forcément le cas. D'autres facteurs expliquent les différences d'usage des dispositifs comme l'usage préalable d'environnements ou la perception de la plus-value apportée par ceux-ci (Mahdizadeh *et al.*, 2007). Pour notre part, nous considérons que, pour faciliter la mise en place des SACI, il faut agir conjointement sur l'expérience professionnelle des enseignants dans leur activité d'encadrement et dans leur activité de conception. Ainsi, nous proposons de favoriser les partages et transferts d'expériences d'apprentissage collectif instrumenté, que ce soit en présence ou à distance. Nous proposons aussi de clarifier le processus de conception des SACI et l'articulation des rôles et moyens d'actions des différents enseignants (concepteurs, tuteurs).

Dans la suite de nos travaux, nous nous attacherons à analyser, formaliser et généraliser des outils d'aide à la conception d'activités de type SACI. Pour cela, nous allons (1) élargir notre étude afin de vérifier les tendances que nous avons dégagées, (2) vérifier s'il existe des invariants aux situations de type SACI (e.g. des liens entre les fonctionnalités de collaboration des plates-formes, les scénarios pédagogiques et les contextes d'apprentissage) et (3) proposer des scénarios types de SACI et des outils d'aide à la conception de ces scénarios se fondant sur des expériences réelles.

**Bibliographie**

Bannon, L.J., "Issues in Computer-Supported Collaborative Learning", C. O'Malley, (Ed.), *Actes de NATO Advanced Workshop on Computer-Supported Collaborative Learning*, Maratea, Italy, septembre 1989.

Béguin, P., « Prendre en compte l'activité de travail pour concevoir », *@ctivités*, vol. 4, n°2, 2007, consulté en ligne le 18/02/2008 http://www.activites.org/v4n2/v4n2.pdf, p. 107-114.

Bourriquen, B., David, J-P., Garrot, E., George, S., Godinet, H., Medélez, E., et Metz, S., « Caractérisation des Situations d'Apprentissage Collectives et Instrumentées dans le supérieur », *8ème Biennale de l'éducation et de la formation*, Lyon, France, 11-14 avril 2006.

Campus SPIRAL pour l'IUT « Techniques de commercialisation » et campus FORSE pour le master 1, http://spiral.univ-lyon1.fr.

Campus FORSE (FOrmation et Ressources en Sciences de l'Éducation), http://www.sciencedu.org/

Caron, P.-A., Ingénierie dirigée par les modèles pour la construction de dispositifs pédagogiques sur des plateformes de formation, Thèse de Doctorat en informatique, Université des Sciences et Technologies de Lille, 2007, 262 p.

Casey, J., Brosnan, K., & Greller, W., Prospects for using learning objects and learning design as staff development tools in higher education. *IADIS International Conference on Cognition and Exploratory Learning in Digital Age (CELDA 2005)*, Porto, Portugal, p. 96-104.

Clot, Y., « Action et connaissance en clinique de l'activité ». *@ctivités*, vol. 1, n°1, 2004, Consulté en ligne le 18/02/2008, http://www.activites.org/v4n2/v1n1.pdf, p. 23-33.

Clot, Y., « Pourquoi et comment s'occuper du développement en clinique de l'activité ? ». *Colloque ARTCO, Symposium International* ,4-5-6 Juillet 2005, Lyon. Consulté en ligne le 18/02/2008 à http://sites.univ-lyon2.fr/artco/telechargement/texte-clot.pdf.

Dillenbourg, P., "What do you mean by 'collaborative learning'?", Dillenbourg P. (Ed.), *Collaborative-learning: Cognitive and Computational Approaches*, Oxford, Elsevier, 1999, p. 1-19.

Garrot E., George S., Prévôt P. (2007) The Development of TE-Cap: an Assistance Environment for Online Tutors. 2nd European Conference on Technology Enhanced Learning (EC-TEL 2007), *Lecture Notes in Computer Science, Springer,* Crete, Greece, p. 481-486.

George, S., Apprentissage collectif à distance - SPLACH : un environnement informatique support d'une pédagogie de projet, Thèse de Doctorat en informatique, Université du Maine, 2001, 354 p.

IMS, IMS Learning Design Information Model – version 1.0. IMS Global Learning Consortium Inc., consulté en octobre 2007
http://www.imsglobal.org/learningdesign/index.html

ISLS, *International Society of the Learning Sciences, CSCL 2007 conference homepage,* http://www.isls.org/cscl2007/index.html

Jermann, P., Soller, A., Muehlenbrock, M., "From mirroring to guiding: A review of state of the art technology for supporting collaborative learning", *European Conference on Computer-Supported Collaborative Learning* (Euro-CSCL 2001), Maastricht, Netherlands, 22-24 mars 2001, p. 324-331.

Koper R., From change to renewal: Educational technology foundations of electronic learning environments. Report from the Open University of the Netherlands, 2006, 50 p. Consulté le 24/10/2007 à http://dspace.ou.nl/bitstream/1820/38/2/koper-inaugural-address-eng.pdf,

Mackay, W., Fayard, A.-L., « Radicalement nouveau et néanmoins familier : les strips papiers revus par la réalité augmentée », *Actes des journées IHM Interaction Homme-Machine*, Poitiers, France, septembre 1997, p. 105-112.

Mahdizadeh, H., Biemans H., Mulder M., "Determining factors of the use of e-learning environments by university teachers", *Computers & Education* (2007), doi:10.1016/j.compedu.2007.04.004 , in press.

McPherson, M., & Nunes, M. B., "The role of tutors as an integral part of online learning support" *European Journal of Open, Distance and E-Learning (EURODL)*, vol. 1, consulté le 24/10/2007, http://www.eurodl.org/materials/contrib/2004/Maggie_MsP.html, 2004

Michel, C., Garrot, E., George, S., (2007a) "Instrumented Collective Learning Situations (ICLS) : the gap between theoretical research and observed practices", *Actes de la 18ème conference internationale sur "Society for Information Technology and Teacher Education" (SITE 2007)*, San Antonio, Texas, EU, 26-30 mars 2007.

Michel, C., Garrot, E., George, S., (2007b) "Situations d'apprentissage collectives instrumentées : étude de pratiques dans l'enseignement supérieur ». *Actes de la Conférence EIAH 2007 - 3eme Conférence en Environnement Informatique pour l'Apprentissage Humain (EIAH 2007)*, Lausanne, Suisse, 27-29 Juin 2007, p.473-478.

Ollagnier, E., « Les exigences de normalisation pour les pratiques de la formation d'adultes en Suisse : réponse des acteurs et analyse critique ». *Congrès self 2004 Ateliers II Les métiers de la formation face aux normes,*2004, p. 485- 492

Perrin, J., *Les transferts de technologie*. Paris, La découverte, 1983.

Rabardel, P., *Les Hommes et les technologies. Approche cognitive des instruments contemporains*, Armand Colin, 1995.

Rogers, E. M., *Diffusion of innovations (5th ed.)*. New York, The Free Press, 2003

West, R. E., Waddoups, G., Graham, C. R., "Understanding the experiences of instructors as they adopt a course management system". *Educational Technology Research and Development*, Springer Boston, 2006, p.1-26.